# Hot ion implantation to create dense NV centre ensembles in diamond

Midrel Wilfried Ngandeu Ngambou[1], Pauline Perrin[2], Ionut Balasa[2], Alexey Tiranov[2], Ovidiu Brinza[1], Fabien Bénédic[1], Justine Renaud[3], Morgan Reveillard[3], Jérémie Silvent[3], Philippe Goldner[2], Jocelyn Achard[1] and Alexandre Tallaire[2]

[1] LSPM, CNRS, Université Sorbonne Paris Nord, 99 avenue JB clément 93460, Villetaneuse, France

[2] IRCP, CNRS, PSL Research University, 11 rue Pierre et Marie Curie, 75005, Paris, France

[3] Orsay Physics, ZAC Saint Charles, 95 avenue des Monts Auréliens, 13710 Fuveau, France

E-mail: midrel.ngandeu@lspm.cnrs.fr; alexandre.tallaire@chimieparistech.psl.eu

Abstract:

Creating dense and shallow nitrogen-vacancy (NV) ensembles with good spin properties, is a prerequisite for developing diamond-based quantum sensors exhibiting better performance. Ion implantation is a key enabling tool for precisely controlling spatial localisation and density of NV colour centres in diamond. However, it suffers from a low creation yield, while higher ion fluences significantly damage the crystal lattice. In this work, we realize $N_2^+$ ion implantation in the 30-40 keV range at high temperatures. At 800 °C, NV's ensemble photoluminescence emission is three to four times higher than room temperature implanted films, while narrow electron spin resonance linewidths of 1.5 MHz, comparable to well-established implantation techniques, are obtained. In addition, we found that ion fluences above $2\times10^{14}$ ions/cm² can be used without graphitization of the diamond film, in contrast to room temperature implantation. This study opens promising perspectives in optimizing diamond films with implanted NV ensembles that could be integrated into quantum sensing devices.

Currently, research groups and industries worldwide are devoting major efforts to exploiting the quantum properties of matter at the nanoscale, as a way to push further the limits and performance of conventional devices and systems. These progresses are set to revolutionize applications such as quantum sensing, quantum computing or cryptography with potential impact in health and medicine, defence, electronics or information processing[1,2]. These activities have generated a thirst for new materials, in particular solid-state systems in which nanoscale defects or impurities are introduced and manipulated on-demand with long coherence times. Diamond is a widely studied host material offering a dense carbon lattice, a low spin bath and a plethora of colour centres[3]. Among them, the negatively charged nitrogen-vacancy ($NV^-$) centre with its photoluminescence (PL) emission at 637 nm is outstanding[4]. It offers excellent spin properties, including millisecond coherence times at room temperature and the possibility to read-out and control spin states using optical and microwave fields. Diamond-based magnetometers[5] are, in particular, starting to be commercially deployed. In this context, optimization of the synthesis of diamond films to fulfil the desired characteristics for quantum sensing is of major interest.

For this purpose, $NV^-$ doped diamonds should ideally possess a low amount of defects, strain, surface damage or paramagnetic impurities that can compromise the coherence time or stability of the negative charge state of the defect. Dense ensembles of NV centres (several tens of ppb) are also usually preferred to increase the number of spins in the sensing volume of the devices, thus improving contrast and sensitivity[6]. For most applications, spatial localization of NVs is also necessary to facilitate their coupling to cavities or nanostructures[7]. For example, in wide-field imaging magnetometry, highly NV-doped nanometre-thin layers should be located near the surface to precisely control the distance between the interacting spins and the sample to be measured[8].

NVs can be introduced during the growth of diamonds via chemical vapour deposition (CVD), by adding a nitrogen-containing gas to the mixture ($N_2$ or $N_2O$, for example)[9]. However, achieving a high spatial localisation is particularly difficult using in-situ doping. Nitrogen ion implantation has thus been widely explored in high-purity diamonds, because it allows controlling depth by setting the implantation energy[10,11] or lateral positioning with a focused ion beam (FIB)[12–14]. Typical ion fluences range from $10^8$ to $10^{15}$ ions/cm² to obtain isolated or dense NV ensembles (several hundreds of ppb). Beyond a certain dose though, when the number of vacancies created by the ion beam is above a threshold of about $10^{22}$ V/cm[3,15], the diamond lattice becomes permanently graphitized and NV properties are degraded. Ion implantation is generally carried out at room temperature or low temperature[16]. However it needs to be followed by an annealing step at typically 800°C [10,12,17,18] to reduce damage and to allow for vacancies to diffuse and combine with nitrogen atoms.

In this work, we obtain high density NV ensembles by using a novel implantation technique at high temperatures (600-900 °C), compatible with sub-micron spatial resolution. Out results pave a clear path towards realisation of a new generation of quantum sensors based on dense NV ensembles, directly useful for a broad range of applications.

Nine diamond films with a 50-100 µm thickness were epitaxially grown onto high pressure high temperature (HPHT) type *Ib* diamond substrates with a homemade plasma-assisted CVD reactor[19]. The films were grown undoped, i.e., without intentional nitrogen addition as confirmed by the lack of any detectable NV emission in PL. After growth, their surface was hydrogen-terminated with a low-power $H_2$ plasma since this treatment favours NV creation efficiency[20]. Implantation of nitrogen was then performed using two experimental set-ups. The first is an internally developed implantation tool that uses a 40 keV $N_2^+$ ion beam source, described in a previous publication[20]. Depth penetration is estimated at 27 nm for this energy (20 keV per N atom) using the free software SRIM[21]. The beam has a relatively large diameter of about 3 mm and allows beam currents of 100 nA or more as measured by a Faraday cup. Hence, it is a simple and robust way to fully implant the entire diamond surface and create uniform layers of NV ensembles. The second experimental setup reaches accurately controlled fluences together with high spatial localization. It consists of a 30 keV $N_2^+$ FIB from *Orsay Physics*. In this system, the beam is mass-filtered to ensure that only the desired ions are implanted, and it is focused down to a 100 nm-size leading to an implantation depth of around 17 nm. We used typical currents of 19 pA for irradiation times of 3 s in 30 µs size spots, to create NVs at a comparable fluence as the previous set-up. Both systems are equipped with a heated stage reaching temperatures of up to 900 °C during implantation. After treatment, all the samples were annealed at 800 °C for 2h to create NVs by vacancy diffusion.

The ion implantation conditions are summarized in Table I. Samples 1 to 4 were dedicated to studying the ion implantation temperature influence. With samples 3, 5 and 6, we checked the effect of increasing fluences on PL emission. Sample 7 was used to confirm the compatibility of hot implantation with spatial localization in small spots created with the FIB. Finally, two more samples (8 and 9) were implanted with higher doses at room and high temperatures to check for possible surface graphitization in 1 mm size spots.

| Sample number | Implantation tool/ pattern | Ion energy (keV) | Fluence (ions.cm$^{-2}$) | Implantation temperature (°C) | Post-implantation annealing |
|---|---|---|---|---|---|
| **1** | **Large beam** | **40** | **$2\times10^{13}$** | **RT** | **800°C/2h** |
| **2** | **Large beam** | **40** | **$2\times10^{13}$** | **600** | **800°C/2h** |
| **3** | **Large beam** | **40** | **$2\times10^{13}$** | **800** | **800°C/2h** |
| **4** | **Large beam** | **40** | **$2\times10^{13}$** | **900** | **800°C/2h** |
| 5 | Large beam | 40 | $6\times10^{13}$ | 800 | 800°C/2h |
| 6 | Large beam | 40 | $1.2\times10^{14}$ | 800 | 800°C/2h |
| 7 | FIB | 30 | $1\times10^{14}$ | RT | 800°C/1h *in-situ* |
| 7 | FIB | 30 | $1\times10^{14}$ | 800 | 800°C/1h *in-situ* |
| 8 | Large beam (1 mm spot) | 40 | $3.1\times10^{14}$ | RT | 800°C/2h |
| 9 | Large beam (1 mm spot) | 40 | $2.2\times10^{14}$ | 800 | 800°C/2h |

*Table I. Diamond samples and nitrogen ($N_2^+$) ion implantation conditions used in this study.*

Photoluminescence was measured using a *Renishaw InVia* Raman spectrometer with a 532 nm excitation laser line. The samples were mounted onto a *Linkam* cooled stage with liquid nitrogen circulation to reach a base temperature of about 80 K. The measurements were acquired with 1800 grooves-grating in a confocal configuration and a 50 × objective. PL images were also acquired using a *DiamondView*$^{TM}$ equipment with near-band edge UV excitation. Optically Detected Magnetic Resonance (ODMR) measurements were performed with a confocal microscopy setup to access the spin properties and measure the coherence times of implanted samples. A high numerical aperture objective (NA = 0.95) was used to focus the excitation green laser (532 nm) of 100 µW onto the sample and collect PL from the $NV^-$ centres. The signal was spatially filtered by a 50 µm-pinhole and finally recorded by a single photon counter detector (*Laser Components COUNT10C*). Additionally, a microwave (MW) field with a tuneable frequency was generated and applied with a 100 µm diameter wire placed near the sample's surface. A permanent magnet was used to lift the degeneracy of the $m_s = \pm 1$ spin sublevels and completely separate lines pertaining to different NV orientations by the Zeeman effect.

Fig. 1a and 1b show the low-temperature PL results following nitrogen implantation and post-annealing at various temperatures and fluences. The emission from $NV^0$ and $NV^-$ centres with zero phonon lines (ZPL) at 575 and 638 nm, respectively, and their phonon sidebands are clearly observed. Their intensities are stronger than the diamond Raman peak, indicating relatively high-density ensembles. We estimate the $NV^-$ concentration to be about 40-200 ppb by comparing the PL intensity to that measured in bulk doped diamond films of calibrated concentrations. Strikingly, we observe an increase of both $NV^-$ and $NV^0$ emission with increasing implantation temperature, reaching a plateau at around 800 °C (red squares of Fig. 1c and 1d) and up to a factor of 3 enhancement. Using higher ion fluences, however, does not directly translate into higher NV emission probably because of saturation of NV creation yield or quenching of PL at higher nitrogen concentrations. We also stress out that precise control of the beam current is difficult to achieve in this set-up, which may lead to inaccuracies that we estimate to be about 10-20 %. Maximum NV emission is obtained for $2\times10^{13}$ ions/cm² implanted at 800 °C and followed by a 2 h annealing step.

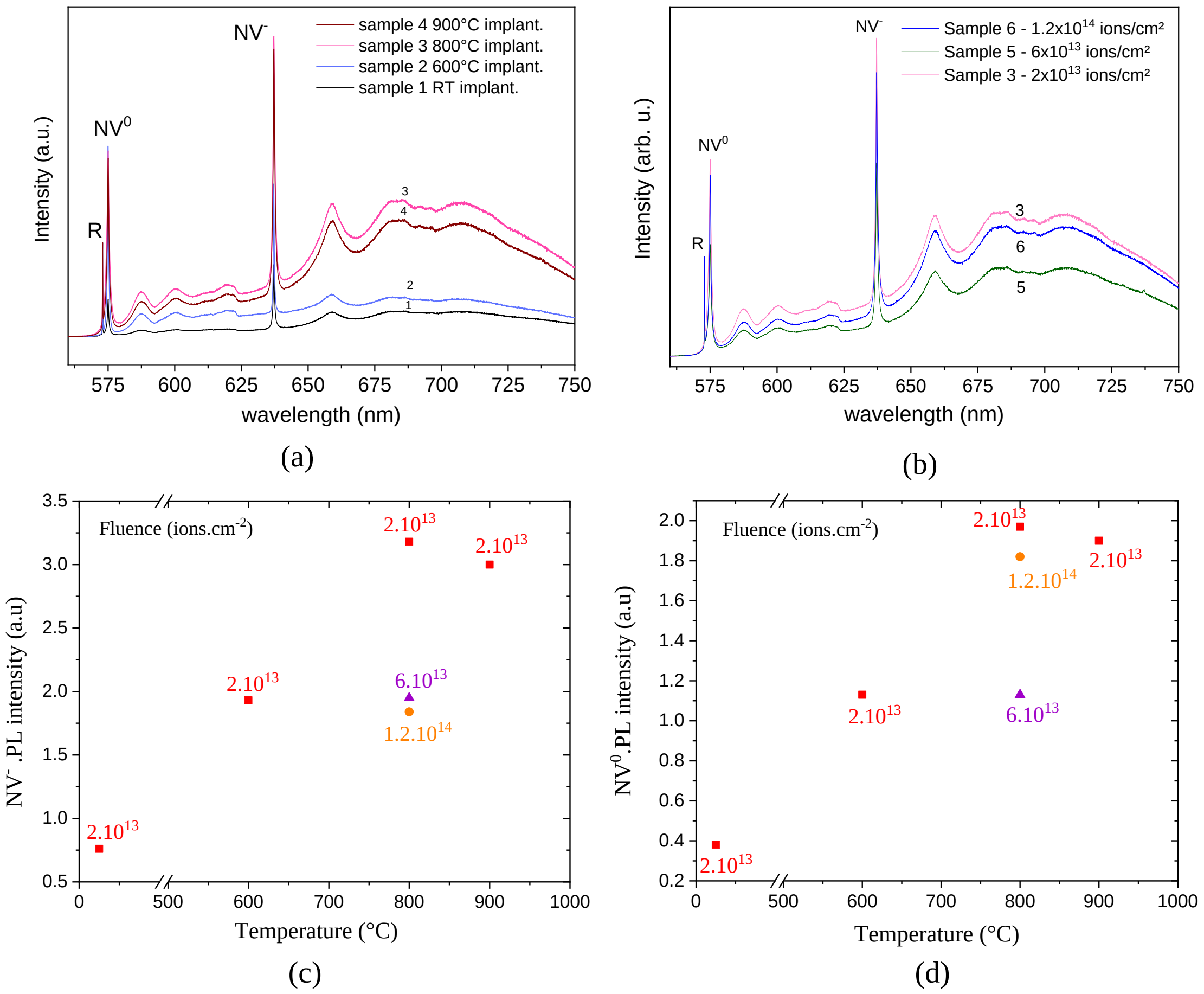

*FIG. 1. PL spectra acquired at 80 K as a function of implantation temperature (a) or ion implantation dose at 800 °C implantation temperature (b). The intensity of $NV^-$ and $NV^0$ emissions are reported in (c) and (d), respectively. All spectra were normalized to the intensity of the diamond Raman peak labelled as R.*

Implanted NVs' spin properties were then assessed by ODMR with continuous excitation. The results are presented in Fig. 2 for room-temperature and 800°C implantation before and after the annealing post-treatment (samples 1 and 3). Electron spin resonance occurs due to population transfer from the bright $m_s = 0$ to the darker $m_s = \pm 1$ spin state at a resonant MW field, which depends on the projection of the applied magnetic field along the NV axis. In the resonance condition, a reduction in PL is observed with a moderate contrast of 0.4 %. Hyperfine coupling to the $^{14}N$ nuclear spin of the NV centre further splits the resonance into 3 lines[6]. By fitting the resulting spectra with 3 Lorentzian functions, the FWHM of one of the lines can be measured in the regime of low applied MW power (0 dBm) to limit power-induced broadening[22]. In this case, the FWHM linewidth (Δν) allows evaluating the lower bound of the dephasing time $T_2$* through the relation:

$$\Delta\nu = \frac{2\sqrt{(ln2)}}{\pi T_2^*} \qquad (1)$$

The results are reported in Fig. 2d. Linewidths in the 1.5 to 1.7 MHz range (equivalent to $T_2$* around 0.3-0.4 μs) were measured which represents a good value for an implanted layer with a high dose of $2\times10^{13}$ ions/cm² [14,23,24]. We also note a slight improvement of ODMR linewidth for the high-temperature implanted sample after annealing compared to the room-temperature implanted one. Since the total amount of nitrogen remains the same, we did not expect a drastic improvement in linewidth because spin properties are limited mainly by the presence of substitutional nitrogen in this doping regime. Nevertheless, this result indicates that hot implantation, besides leading to a better N to NV conversion, also allows for preserving narrow ODMR lines, a clear advantage for developing quantum sensors with improved sensitivity.

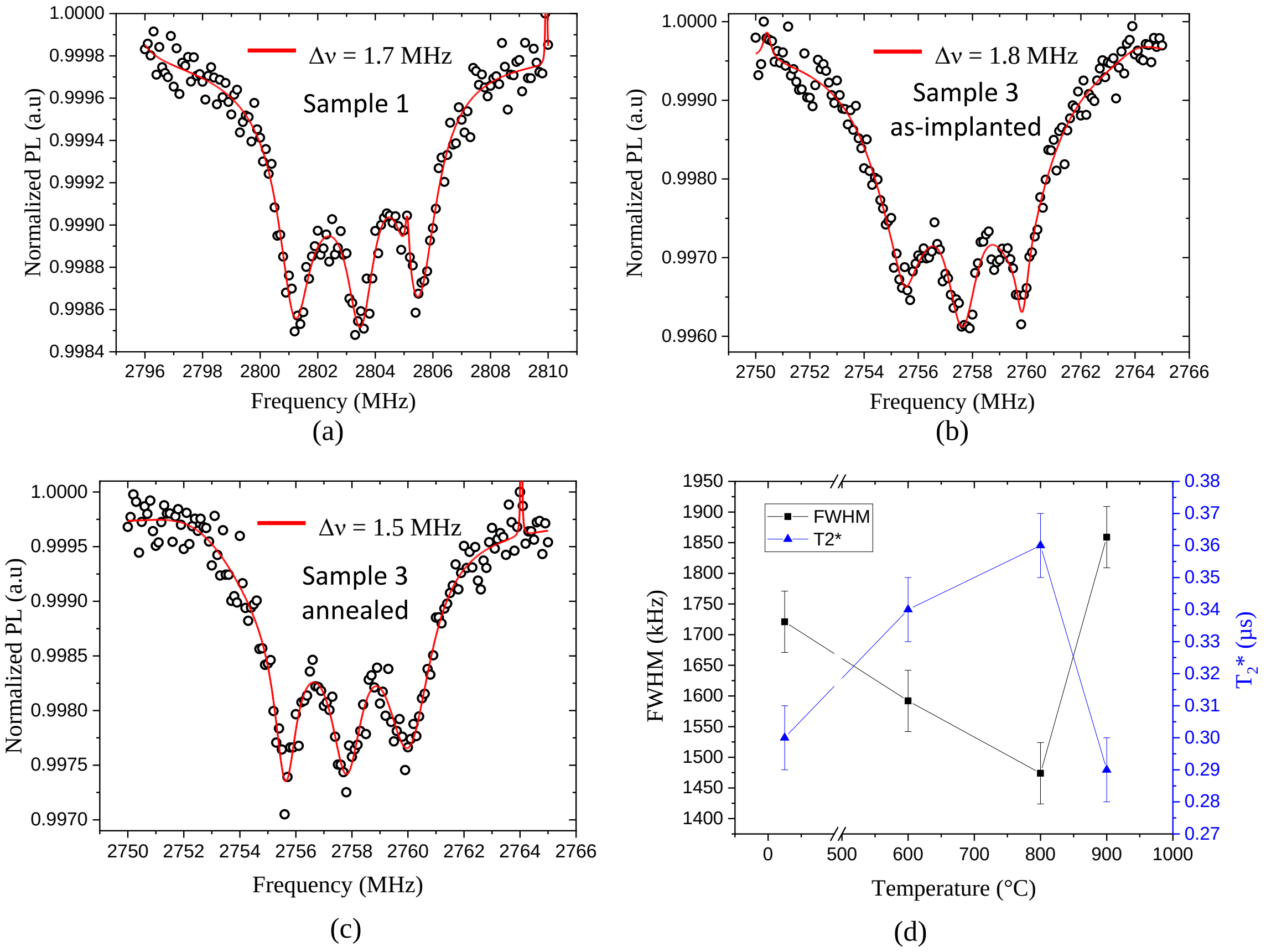

*FIG. 2. ODMR spectra of sample 1, implanted at room temperature (a) and sample 3 implanted at 800 °C before (b) and after post-annealing (c) are shown. The measured linewidth of one of the hyperfine lines is given for each spectrum and the results are further plotted in (d) for all the studied implantation temperatures. $T_2^*$ is also estimated from this linewidth.*

To further confirm the compatibility of this high-temperature implantation approach with accurate NV spatial localization, we repeated the experiment with the FIB setup under highly controlled fluences[12]. To facilitate observation, two rings were first milled using $Ar^+$ at 30 keV. This was immediately followed by ion implantation either at room temperature or 800 °C inside each ring, using $N_2^+$ at 30 keV and a $10^{14}$ ions/cm² fluence. The sample was then in-situ annealed for 1 h at 800 °C. PL maps at around 638 nm (Fig. 3a and 3b) show that NV centres were successfully created and localized inside the rings in both regions. A comparison of PL intensity, acquired in strictly identical conditions (Fig. 3c), confirms a clear improvement for the hot implanted sample, this time by about a factor of 4.

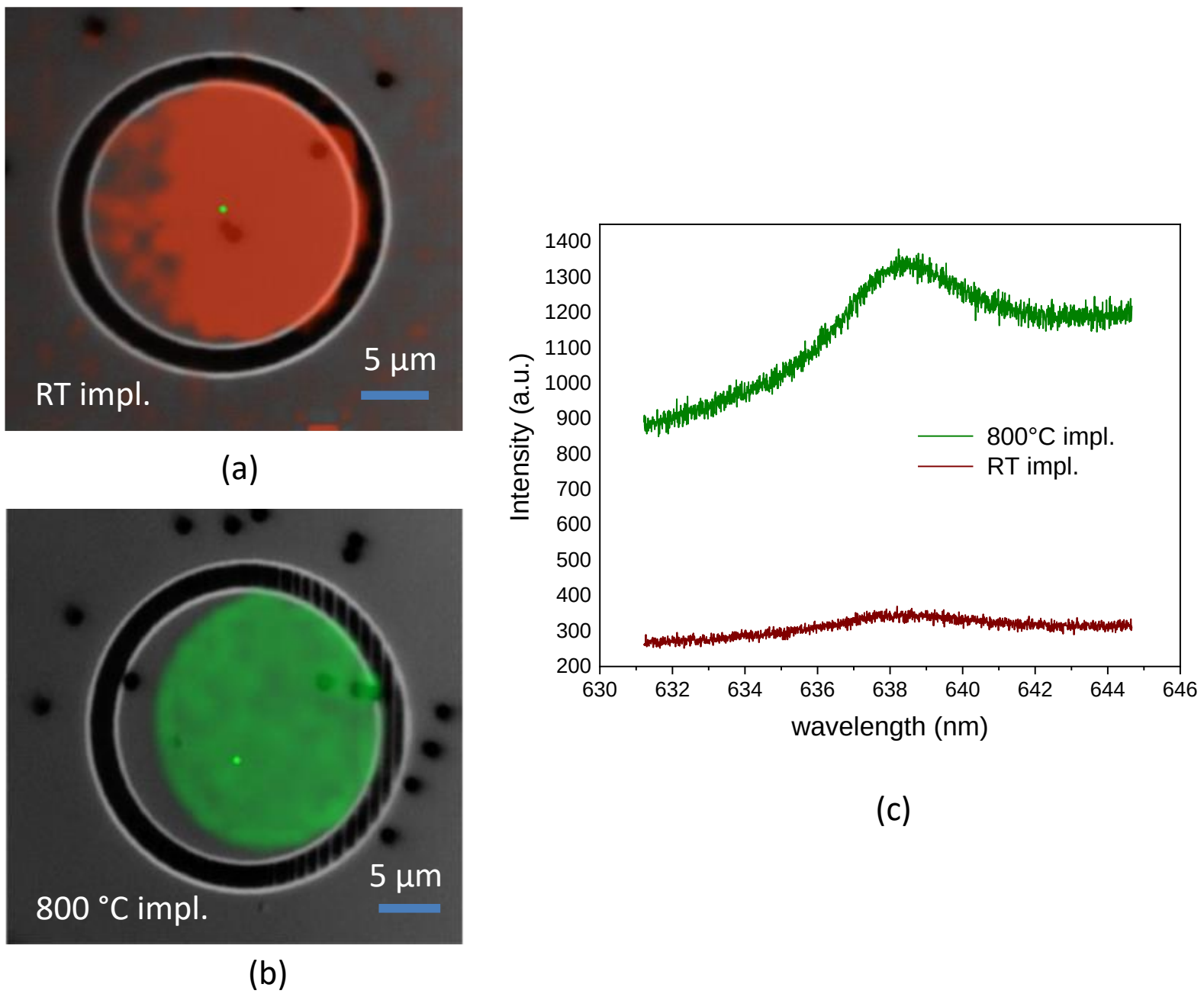

*FIG. 3. Photoluminescence maps acquired for sample 7 at a wavelength of 638 nm from two spots implanted with a FIB at room temperature in red (a) or 800 °C in green (b). PL spectra in this spectral region are plotted in (c) to compare their intensity.*

Finally, to get more insight into the improvement mechanism, we tested how graphitization occurs by implanting two samples (n°8 and 9) at higher fluences (> $2\times10^{14}$ ions/cm²) either at room temperature or at 800 °C in the first setup. A physical metallic mask with a 1 mm aperture was placed between the diamond and the beam, resulting in a large implantation spot. The *DiamondView*™ PL images are acquired under UV light at various processing steps (Fig. 4a and 4b). Bright green luminescence from the type-*Ib* HPHT substrate is observed, while the pure CVD layer on top is mostly transparent. Strikingly, the room-temperature implanted diamond exhibits a dark spot soon after implantation, which only partially converts to NV centres after annealing. On the contrary, the diamond implanted at 800 °C does not show any graphitization but red luminescence from NV centres that gets brighter after the post-treatment annealing. This is further confirmed by the Raman analysis carried out after annealing. While the G-band related to graphite is observed for sample 8, only the diamond peak appears for sample 9. From this we conclude that hot implantation allows much higher fluences to be used before reaching the graphitization threshold of diamond. Implanting at a temperature at which vacancies are mobile[10] (> 700 °C) probably limits defect formation and damage. Interstitial-vacancy pairs created by the beam quickly recombine, avoiding the formation of vacancy clusters[25,26] (such as $V_2$) that can lead to a dramatic degradation of the diamond lattice or graphitization. Introduced nitrogen atoms also quickly combine with adjacent vacancies produced by the beam, thus limiting the competition between NVs and other defects formation during the post-annealing.

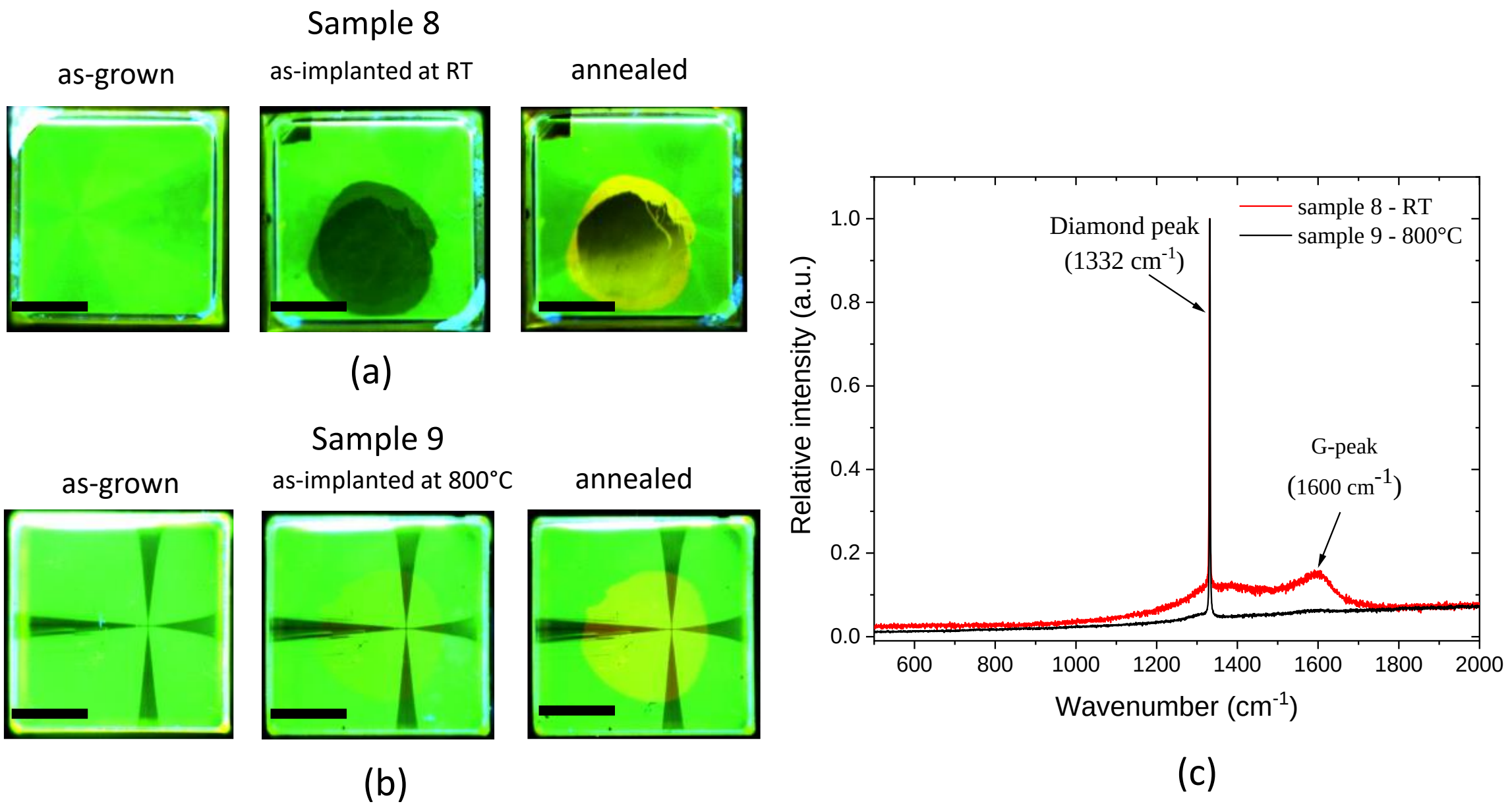

*FIG. 4. DiamondView$^{TM}$ PL images obtained for the CVD films grown on Ib HPHT substrates: sample 8, room-temperature implanted (a) and sample 9, 800 °C implanted (b). Images were taken using the same acquisition conditions after each treatment step. The black scale bar is 1 mm. (c) The Raman spectra acquired in the implanted region of each sample are shown, and both the diamond Raman and graphite G-peak are identified.*

In summary, we created dense NV ensembles of a few tens of ppb in pure CVD diamond films by using hot nitrogen ion implantation. At 800 °C, NV PL emission is increased by a factor of 3 to 4 while good spin properties are preserved (with ODMR linewidths of about 1.5 MHz). The approach is also compatible with spatial localisation using a FIB system. Compared to standard room temperature implantation, hot implantation allows introducing higher amounts of nitrogen before dramatic graphitization of the diamond film occurs. Finding optimised ion implantation conditions is key to control the density and properties of NVs in diamond and is thus a crucial step towards the development of diamond-based quantum sensors. This approach could further be extended to the creation of other defects in diamond such as group IV colour centres (SiV, GeV etc.) or the recently discovered ST1 centre that involves oxygen atoms[27] as well as other material platforms including defects in SiC[28].

## Acknowledgements

The ANR Equipex+ project (*e-diamant*) is thanked for financially supporting the FIB and Raman analysis platforms. The ANR projects *Diamond-NMR* and *Trampoline*, the *Quantera* project *MAESTRO* and CNRS Prime 80 grant (*MATHYQ*) are gratefully acknowledged for funding. This project has received funding from the European Research Council (ERC) under the European Union's Horizon 2020 research and innovation programme (grant agreement No 101019234, *RareDiamond*). The authors would like to thank Jean-François Roch from ENS Paris-Saclay for discussions and scientific support in this study.